\documentclass[a4paper,aps,prd,superscriptaddress,preprint]{revtex4}
\usepackage[utf8]{inputenc}
\usepackage{color}
\usepackage{textcomp}
\usepackage{amsmath}
\usepackage{amssymb}
\usepackage{graphicx}

\makeatletter

\pdfpageheight\paperheight
\pdfpagewidth\paperwidth

\@ifundefined{textcolor}{}
{%
 \definecolor{BLACK}{gray}{0}
 \definecolor{WHITE}{gray}{1}
 \definecolor{RED}{rgb}{1,0,0}
 \definecolor{GREEN}{rgb}{0,1,0}
 \definecolor{BLUE}{rgb}{0,0,1}
 \definecolor{CYAN}{cmyk}{1,0,0,0}
 \definecolor{MAGENTA}{cmyk}{0,1,0,0}
 \definecolor{YELLOW}{cmyk}{0,0,1,0}
}

\usepackage{bm}
\usepackage{indentfirst}
\usepackage{float}
\usepackage{caption}
\usepackage{psfrag}
\usepackage{epstopdf}
\usepackage{diagbox}
\usepackage{orcidlink}

\@ifundefined{showcaptionsetup}{}{%
 \PassOptionsToPackage{caption=false}{subfig}}
\usepackage{subfig}
\makeatother

\begin{document}
\title{Hunting for extra dimensions in black hole shadows}
\author{A. S. Lemos \orcidlink{0000-0002-3940-0779}}
\email{adiel@ufersa.edu.br}

\affiliation{Departamento de Ci\^encias Exatas e Tecnologia da Informa\c c\~ao, Universidade
Federal Rural do Semi-Árido, 59515-000 Angicos, Rio Grande do Norte,
Brazil}
\affiliation{Departamento de F\'isica, Universidade Federal de Campina Grande, Caixa
Postal 10071, 58429-900 Campina Grande, Para\'iba, Brazil}
\author{J. A. V. Campos \orcidlink{0000-0002-4252-2451}}
\email{joseandrecampos@gmail.com}

\affiliation{Departamento de F\'isica, Universidade Federal de Campina Grande, Caixa
Postal 10071, 58429-900 Campina Grande, Para\'iba, Brazil}
\author{F. A. Brito \orcidlink{0000-0001-9465-6868}}
\email{fabrito@df.ufcg.edu.br}

\affiliation{Departamento de F\'isica, Universidade Federal de Campina Grande, Caixa
Postal 10071, 58429-900 Campina Grande, Para\'iba, Brazil}
\affiliation{Departamento de F\'isica, Universidade Federal da Para\'iba, Caixa Postal
5008, 58051-970 Jo\~ao Pessoa, Para\'iba, Brazil}
\begin{abstract}
Observational data of the Sagittarius A{*} (Sgr A{*}) shadow released
by the Event Horizon Telescope are used to investigate eventual
deviations in the black hole shadow radius, aiming to seek physics
beyond the Standard Model coming from extra-dimensional theory.
We consider the braneworld scenario described by the Randall-Sundrum
model and determine the black hole shadow radius correction owing
to the higher dimension. From data of the shadow radius in units of
black hole mass determined by Keck- and Very Large Telescope Interferometer (VLTI)-based estimates, we imposed restrictions
on the deviation obtained, and one sets an upper limit to the curvature
radius of anti-de Sitter ($\mathrm{AdS_{5}}$) spacetime $\ell_{\left(\text{Keck}\right)}\lesssim3.8\times10^{-2}\,\mathrm{AU}$
and $\ell_{\left(\text{VLTI}\right)}\lesssim1.6\times10^{-2}\,\mathrm{AU}$
(at $68\%$ confidence level), respectively.
\end{abstract}
\maketitle

\section{Introduction}

Over the past century, attempts have been made to investigate the
geometrical structure of the Universe from the proposal of the existence
of extra dimensions \citep{Nordstrom:1914ejq}. Following the development
of general relativity (GR), a theoretical extra-dimensional model was built,
which aimed to unify the electromagnetic and gravitational fields
\citep{Kaluza:1921tu,Klein:1926tv}. Recently, the modern braneworld
theories have renewed interest in higher dimensional theories \citep{Antoniadis:1998ig,Arkani-Hamed:1998jmv,Arkani-Hamed:1998sfv,Randall:1999ee,Randall:1999vf}.
These proposed models arise in the unification schemes, where one
expected that the unification of gravitational and gauge forces would
occur near the $\mathrm{TeV}$ scale \citep{Antoniadis:1998ig,Arkani-Hamed:1998jmv,Arkani-Hamed:1998sfv}.
In such a case, the gauge interactions, and therefore the ordinary matter,
get confined to a four-dimensional hypersurface (three-brane), whereas
gravity lives in a higher-dimensional spacetime known as bulk \citep{Randall:1999vf,Randall:1999ee}.
Thus, gravity would be diluted in the higher-dimensional space, spreading
throughout the bulk, explaining, therefore, the observed weakness
of gravity compared to other fundamental interactions of nature \citep{Antoniadis:1998ig,Arkani-Hamed:1998jmv,Arkani-Hamed:1998sfv,Randall:1999ee,Randall:1999vf}.
Consequently, this is a promising scenario to address the hierarchy
problem that deals with the discrepancy between the gravitational
and electroweak fundamental energy scales without relying on low-energy
supersymmetry or technicolor \citep{Antoniadis:1998ig,Arkani-Hamed:1998jmv,Arkani-Hamed:1998sfv}.

In this framework, although the Standard Model (SM) fields are trapped
on our four-dimensional manifold, events of high-energy collisions can,
for instance, eventually cause the SM particles to be kicked off into
the new dimensions \citep{ParticleDataGroup:2022pth}. Thus, collider
experiments have been employed to seek evidence of large extra
dimensions in the Arkani-Hamed-Dimopoulos-Dvali (ADD) model, and it was
found that the most stringent limits on the Planck mass of higher-dimensional
spacetime come from the ATLAS (CMS) missing energy experiment, giving
$M_{D}>5.9–11.2\,\mathrm{TeV}$ \citep{ATLAS:2021kxv}.
In its turn, analysis from astrophysical and cosmological origin,
such as light Kaluza-Klein (KK) graviton production in stars, has also provided
the strongest bounds so that $M_{D}>1700\,(76)\,\mathrm{TeV}$ for
$\delta=2\left(3\right)$ \citep{Hannestad:2003yd}. Further discussions
have attempted to seek traces of extra dimensions in several physical
systems, such as in rotating torsion balance \citep{Adelberger:2009zz,Murata:2014nra}---wherein one tests Newton's law at sub-mm distances---spectroscopy
and atomic physics \citep{Dahia:2015xxa,Dahia:2015bza,Dahia:2017jya,Lemos:2019rhj,Lemos:2021xpi,Luo:2006ad,Zhi-gang:2007swh,Salumbides:2015qwa,Lemos:2022vhw},
and black hole production \citep{CMS:2018ozv,ATLAS:2015yln}.

On the other hand, the model proposed by Randall and Sundrum (RS)
\citep{Randall:1999ee,Randall:1999vf} describes our Universe by postulating
it as a five-dimensional anti-de Sitter ($\mathrm{AdS_{5}}$) spacetime
with a warped geometry. In this scenario, bounds on the $\mathrm{AdS_{5}}$
curvature radius $\ell$ have also been imposed from astrophysical
observations. For instance, the investigation of the persistence of
x-ray binaries containing a black hole in an $\mathrm{AdS_{5}}$ braneworld
shows that $\ell\lesssim10^{-2}\,\mathrm{mm}$ \citep{Emparan:2002jp}.
Furthermore, recent work has also aimed to search for signals of the
extra dimension on events of black hole gravitational waves \citep{Bohra:2023vls,Chrysostomou:2022evl}.
In such a way, the study of Hawking radiation emitted by stellar-mass
black holes found that $\ell\lesssim5\,\mathrm{\mu m}$ from gravitational
wave measurements \citep{McWilliams:2009ym}, while Ref. \citep{Vagnozzi:2019apd}
---analyzing the shadow of M87{*}---imposes constraints on the
$\mathrm{AdS_{5}}$ radius such that $\ell\lesssim170\,\mathrm{AU}$.
The multimessenger cosmology, which considers joint observations
of gravitational waves and electromagnetic counterparts, has been
used to set limits on the radius of curvature of the higher-dimensional
spacetime, $\ell\lesssim0.535\,\mathrm{Mpc}$ \citep{Visinelli:2017bny}.
On its turn, Ref. \citep{Mishra:2021waw} constrains the tidal charge
of a rotating braneworld black hole, whereas in Ref. \citep{Hou2021}
one investigates how the extra dimension affects the shape of the black
hole shadows.

It is known that the black holes only recently have been experimentally probed. Thus, one expect that black hole physics may provide us with significant insights into the structure of spacetime \citep{Chakraborty:2017qve, Dey:2020pth, Hashemi:2019jlt, Anacleto:2023ntm, Cruz:2020emz, Cardoso:2003cj, Konoplya:2011qq, Konoplya:2003ii}, as well as allow us to test fundamental physics \citep{Yagi:2016jml,Goddi:2016qax}. In this context, recent measurements of the Event Horizon Telescope (EHT)
provided images of supermassive black hole shadow at the M87{*} galaxy
center \citep{EventHorizonTelescope:2019dse,EventHorizonTelescope:2019ggy},
in addition to the shadow of compact source Sagittarius A{*} (Sgr
A{*}), a supermassive black hole in the center of the Milky Way \citep{EventHorizonTelescope:2022xqj,EventHorizonTelescope:2022wkp}.
It is noteworthy that, in this scenario, the search for physics beyond
the SM has been based on analysis of the shadow radius and deviation
from the circularity $\Delta C$ of the black hole shadow \citep{vagnozzi2022horizon}.
However, whereas the limit on the circularity deviation has been quoted for the
M87{*} measurements, imposing $\Delta C\lesssim10\%$ \citep{EventHorizonTelescope:2019dse},
such information for Sgr A{*} is not available. Therefore, the approach employed in Ref. \citep{Vagnozzi:2019apd}
cannot be applied to test fundamental physics and to put limits on the $\mathrm{AdS_{5}}$ curvature radius from the circularity
deviations of Sgr A{*} \citep{Vagnozzi2022}. Instead, one must proceed with an investigation based on its shadow size.

Since the extra dimensions have been proposed in modified frameworks of GR, one expects that they might yield signals on the black hole shadow. Therefore, black hole images can be used to test fundamental physics \citep{Yagi:2016jml,Vagnozzi2022}. In this work, we perform a study of the shadow radius of Sgr A* aimed at obtaining new and complementary constraints on the extra-dimensional $\mathrm{AdS_{5}}$ curvature radius from observational data. In this case, we imposed limits on deviations from the GR's predictions arising from a braneworld theory. For this purpose, we will consider
a black hole in the RS background in the weak field regime, described
by the Garriga-Tanaka metric \citep{Garriga:1999yh}. This solution
has been obtained by assuming a classic approach that provides the
linearized braneworld gravity. As we will see, our
analysis allowed us to place the bounds on the extra-dimensional curvature
radius $\ell$ in an alternative way, avoiding the issue of lack of
constraints on the deviation from circularity of the shadow measures
for Sgr A{*} and exploring its inaccurate spin measurements \citep{Banerjee:2022jog,Fragione:2020khu}.

The organization of this paper is as follows: In
Sec. \ref{S2}, by considering the Randall-Sundrum braneworld scenario, we investigate the null geodesic curves aiming
to find the shadow radius of the RSII black hole and discuss the
instability of the null circular orbits from the Lyapunov exponent. In Sec. \ref{S3}, from the EHT observational data of the Sgr A*'s image black hole, we
set the imposed constraints on the $\mathrm{AdS_{5}}$ curvature
radius $\ell$. The concluding remarks and the main results are given
in Sec. \ref{S4}.

\section{Corrected Shadow Radius of the Braneworld Black Hole \label{S2}}

In this section, we will study the null geodesic
of the Garriga-Tanaka (RSII) black hole, aiming to investigate the
influence of the extra dimension on the critical impact and critical
radius parameters. This approach will also allow us to determine the
radius of the shadow of the braneworld black hole. From this investigation,
we intend to find bounds on the curvature radius of the $\mathrm{AdS_{5}}$
spacetime. In this case, we start by considering the Garriga-Tanaka
solution that, at the first-order approximation of $GM$, describes
the metric of a matter distribution with mass $M$ localized in the
brane \citep{Garriga:1999yh},
\begin{equation}
ds^{2}=-f(r)dt^{2}+\dfrac{dr^{2}}{g(r)}+r^{2}d\Omega^{2},\label{el1}
\end{equation}
where, in the weak field limit $\left(r\gg\ell\right)$ from the horizon
of matter after the collapse, we get 
\begin{eqnarray}
f(r)=1-\dfrac{2M}{r}-\dfrac{4M\ell^{2}}{3r^{3}},\quad\text{and}\quad g(r)=1-\dfrac{2M}{r}-\dfrac{2M\ell^{2}}{3r^{3}},
\end{eqnarray}
in which $\ell$ is the curvature radius of $\mathrm{AdS_{5}}$, and
we consider unities $c=G=1$. We can find the geodesics of the metric \eqref{el1}, considering the following
Lagrangian density
\begin{equation}
\mathcal{L}\equiv\frac{1}{2}g_{\mu\nu}\dot{x}^{\mu}\dot{x}^{\nu},
\end{equation}
where the dot denotes the derivative with respect to the affine parameter.
Now, applying the line element \eqref{el1}, we have 
\begin{equation}
2\mathcal{L}=f(r)\dot{t}^{2}-\dfrac{\dot{r}^{2}}{g(r)}-r^{2}\left(\dot{\theta}^{2}+\sin^{2}\theta\dot{\phi}^{2}\right).\label{elidot}
\end{equation}

\begin{figure}
\includegraphics[scale=0.55]{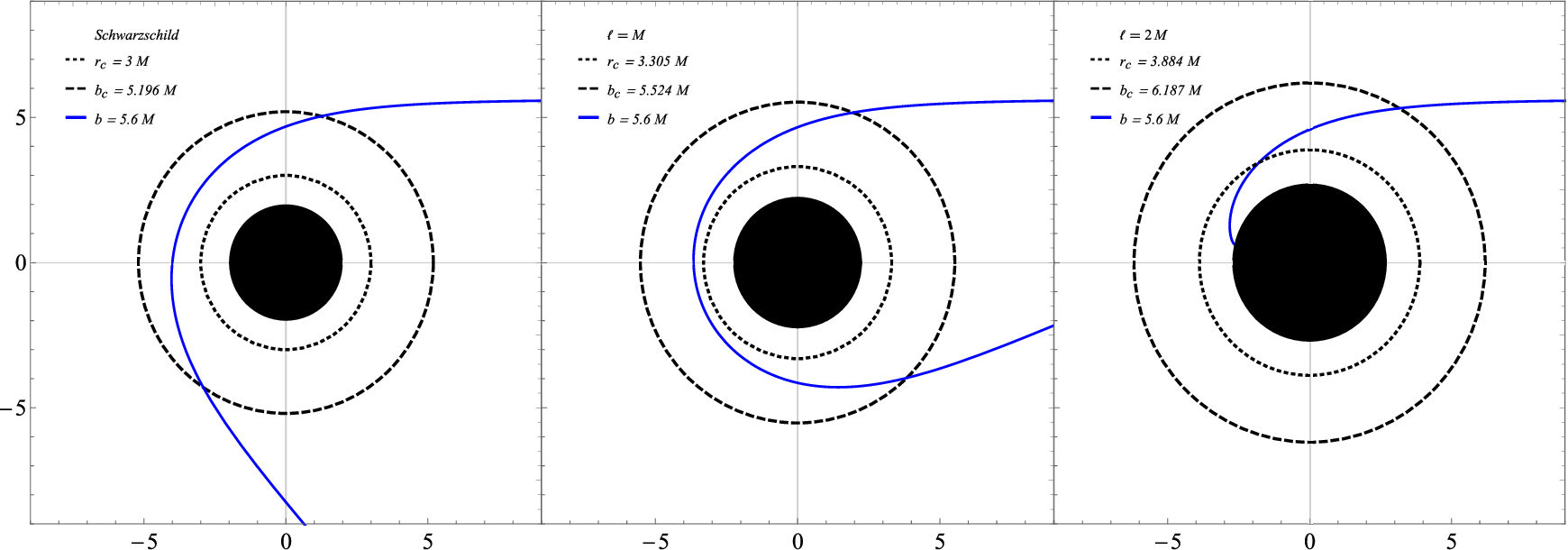} \caption{\label{geod}Null geodesic in polar coordinates for different values
of $\ell$, in which we consider a test beam of light by keeping the
impact parameter, $b=5.5M$, fixed.}
\end{figure}

For the sake of simplicity, at this point, by setting the angle $\theta=\pi/2$,
one can examine the motion in an equatorial plane. Thus, we can find
the equation of motion for the Hamilton-Jacobi equation 
\begin{equation}
E=f(r)\dot{t},\qquad\quad L=r^{2}\dot{\phi},\label{EL}
\end{equation}
where $E$ and $L$ are the constants corresponding to energy and
angular momentum, respectively. For a null geodesic, we have $g_{\mu\nu}\dot{x}^{\mu}\dot{x}^{\nu}=0$,
and we can find the following expression 
\begin{equation}
\dot{r}^{2}+\mathcal{V}(r)=0,\qquad\mathcal{V}(r)=\dfrac{g(r)}{f(r)}E^{2}-g(r)\dfrac{L^{2}}{r^{2}}.\label{eqEner}
\end{equation}
Now, just for convenience, we can introduce a new variable $u=1/r$
to write the orbital equation 
\begin{equation}
\left(\dfrac{du}{d\phi}\right)^{2}=\dfrac{\left(3-6Mu-2M\ell^{2}u^{3}\right)}{\left(3-6Mu-4M\ell^{2}u^{3}\right)b^{2}}-\left(1-2Mu-\dfrac{2M\ell^{2}u^{3}}{3}\right)u^{2},\label{eqD1}
\end{equation}
\begin{equation}
\dfrac{d^{2}u}{d\phi^{2}}=\dfrac{2M\ell^{2}\left(3-4Mu\right)u^{2}}{b^{2}\left(6Mu+4M\ell^{2}u^{3}-3\right)^{2}}-\left(1-3Mu-\dfrac{5M\ell^{2}u^{3}}{3}\right)u.\label{eqD2}
\end{equation}
Here, we have that $b=L/E$ is the impact parameter defined as the
perpendicular distance (measured from infinity) between the geodesic
and the parallel line that passes through the origin. Using the following
critical conditions, (i) $\mathcal{V}(r_{c})=0$ and (ii) $d\mathcal{V}(r)/dr\big|_{r=r_{c}}=0$,
we can obtain the critical impact parameter from a system of equations
with which one relates the impact parameter to the critical radius,
as follows: 
\begin{eqnarray}
b_{c}^{2}=\dfrac{r_{c}^{2}}{f(r_{c})}=\dfrac{3r_{c}^{5}}{3r_{c}^{3}-6Mr_{c}^{2}-4M\ell^{2}}.\label{b_crit}
\end{eqnarray}
Notice that $r_{c}$ corresponds to the radius of photon circular
orbit. On its turn, the critical conditions expressed by (ii) yield
a polynomial equation that is used to determine the critical radius:
\begin{eqnarray}
20\ell^{4}M^{2}+6\ell^{2}M\left(13M-6r_{c}\right)r_{c}^{2}+9r_{c}^{4}\left(6M^{2}-5Mr_{c}+r_{c}^{2}\right)=0.
\end{eqnarray}
We must highlight that, here, we want to study the trajectory of a
ray of light in the spherically symmetric Garriga-Tanaka background.
Specifically, when we analyze the ray in a plane, we observe that
any ray of light that starts at a particular angle $\theta$ must
keep the same angle throughout its path. In Fig. \ref{geod}, we
have found some results for the critical impact parameter from the
numerical solution of the orbit Eqs. \eqref{eqD1} and \eqref{eqD2}.
In this case, we have obtained the geodesic trajectory (blue thick
curve) computed for the impact parameter $b=5.5M$. Therefore, as
a direct influence of the existence of the extra-dimensional parameter
$\ell$, we observed an increase in the absorption section. Furthermore,
in Fig. \ref{geod}, whereas the black disk depicts the limit of
the event horizon, the critical radius for the photon sphere is shown
as an internal dotted circle. In turn, the critical impact parameter
(black hole shadow) is shown by an external dashed circle.

\begin{figure}[!ht]
\includegraphics[scale=0.25]{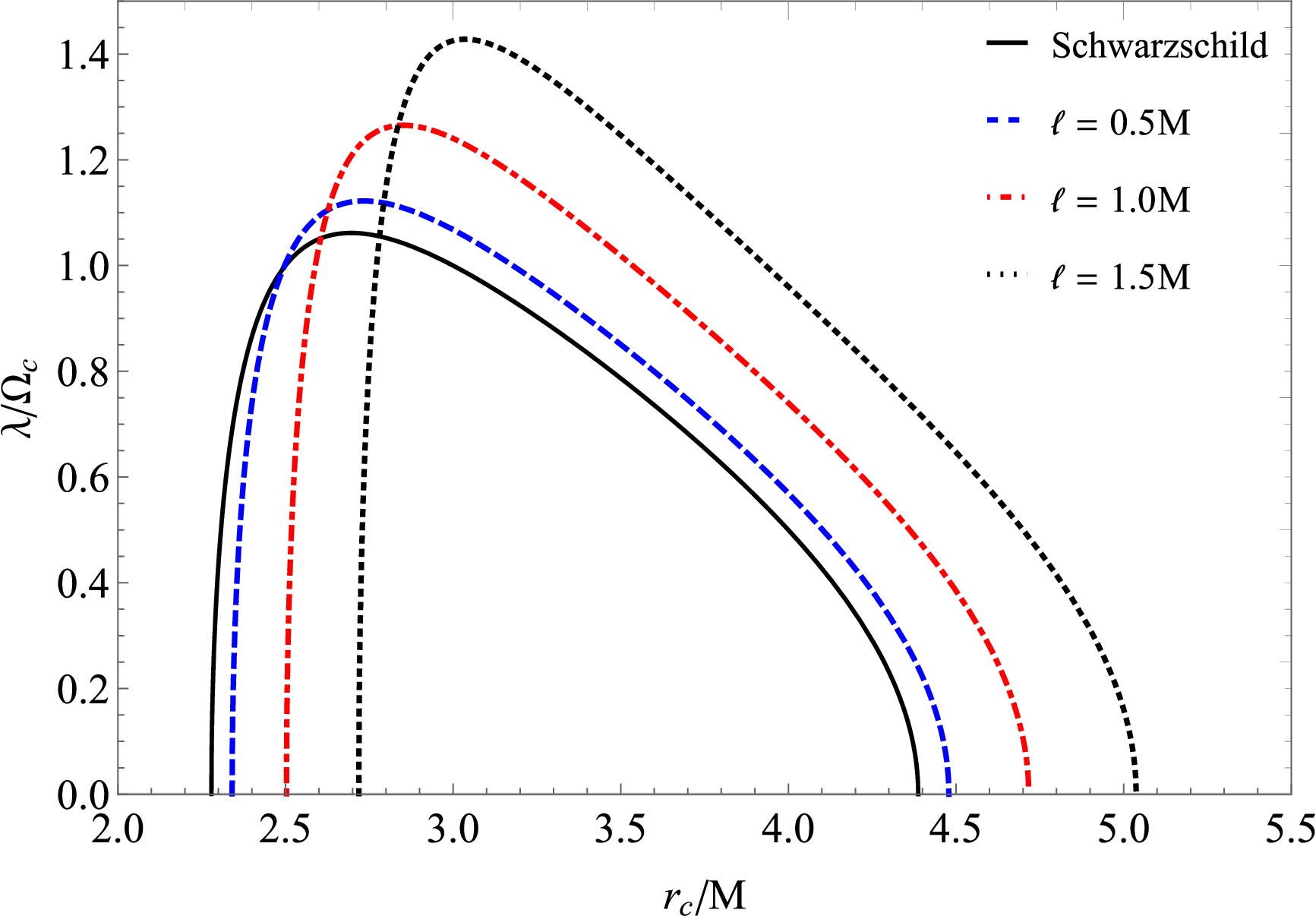} \caption{\label{lyapunov} Lyapunov exponent normalized by the angular velocity
$\left(\lambda/\Omega_{c}\right)$, as function of $r_{c}/M$ for
some values of $\ell$.}
\end{figure}

In this scenario, it is important to verify the instability of null circular orbits. For this, we have two important quantities, namely the Lyapunov exponent $\lambda=\sqrt{\mathcal{V}''(r)/2\dot{t}^{2}}$ that determines the unstable timescale and the angular velocity $\Omega_{c}=\dot{\phi}/\dot{t}$ for the last null circular orbit \citep{cardoso2009geodesic,Anacleto:2021qoe, Campos:2021sff}. The instability exponent is defined by the ratio of the Lyapunov exponent to the angular velocity. In Fig. \ref{lyapunov}, we see the variation
of the instability exponent in relation to the radius $r_{c}$. The
maximum of the instability increases with the increase in the contribution
coming from the extra dimension. On the other hand, the instability
decreases as $r_{c}$ increases. It is worth mentioning that this result is directly sensitive to the geometry, so this analysis is expected to eventually be used to distinguish between Schwarzschild and braneworld black holes.

Finally, one can estimate the size of the radius of the black hole
shadow by using the study of geodesic curves that we might express
via celestial coordinates \citep{vazquez2003strong} 
\begin{eqnarray}
\alpha & = & \lim\limits _{r_{o}\to\infty}\left[-r_{o}^{2}\sin\theta_{o}\dfrac{d\phi}{dr}\Bigr\rvert_{\theta=\theta_{o}}\right],\label{coodcelesta}\\
\beta & = & \lim\limits _{r_{o}\to\infty}\left[r_{o}^{2}\dfrac{d\theta}{dr}\Bigr\rvert_{\theta=\theta_{o}}\right],\label{coodcelestb}
\end{eqnarray}
where $\left(r_{o},\theta_{o}\right)$ is the position of the observer
at infinity. Thus, for an observer on an equatorial plane $\theta_{o}=\pi/2$,
we have the following relationship for the shadow radius 
\begin{equation}
R_{s}\equiv\sqrt{\alpha^{2}+\beta^{2}}=\sqrt{\dfrac{3r_{c}^{5}}{3r_{c}^{3}-6Mr_{c}^{2}-4M\ell^{2}}}.\label{rshw}
\end{equation}
Let us now expand the radius of the shadow up to order $\left(\ell^{2}/M^{2}\right)$
by assuming that the contribution arising from extra dimension is
a small correction to $R_{s}$, i.e., $\ell/M\ll1$, so that 
\begin{equation}
R_{s}\approx3\sqrt{3}M\left[1+\frac{2}{27}\frac{\ell^{2}}{M^{2}}+\mathcal{O}\left(\frac{\ell^{4}}{M^{4}}\right)\right].\label{RS}
\end{equation}
Notice that the extra-dimensional deviation can be
understood as a correction to the Schwarzschild's shadow radius estimate. In the next section, we will find limits on the correction to the shadow radius.

\section{Constraints on the $\mathrm{AdS_{5}}$ curvature radius \label{S3}}

As we saw above, we obtained deviations in the black hole shadow radius
owing to the existence of an extra dimension. On its turn, if we turn
off the extra-dimensional effects ($\ell=0$), we recover the well-known
results for the Schwarzschild background. Although this independent
approach allows us to search for traces of extra dimensions, imposing
constraints on deviations found, once the curvature radius $\ell$
is positively correlated to the shadow radius, and given the low-precision
measurements, thus we must have a weakening in the predicted bounds
compared to sub-mm gravity experiments, for instance.

\begin{figure}
\subfloat[\label{shadow}]{\includegraphics[scale=0.3]{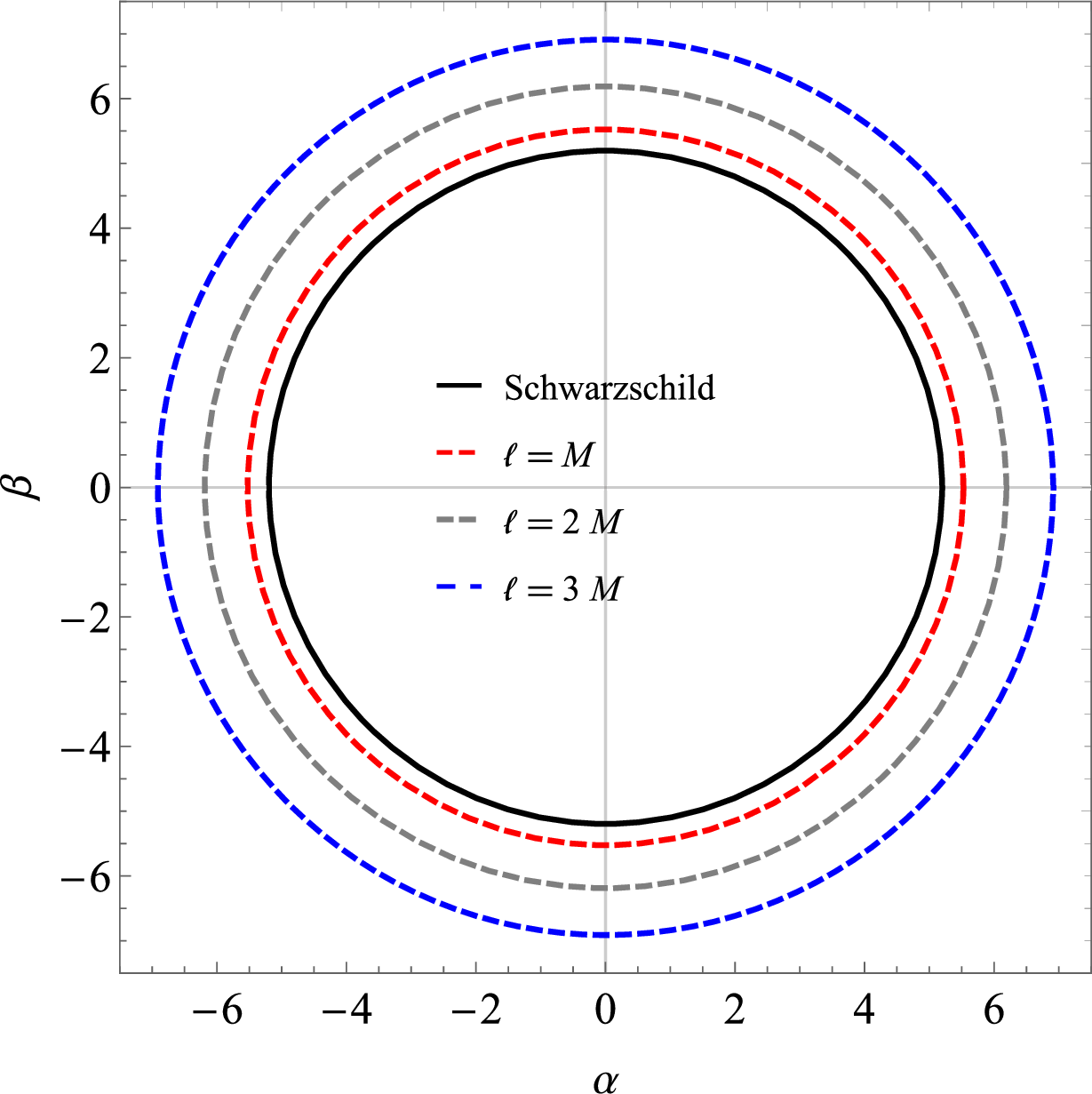}

}\hfill{}\subfloat[\label{Shcomp}]{\includegraphics[scale=0.3]{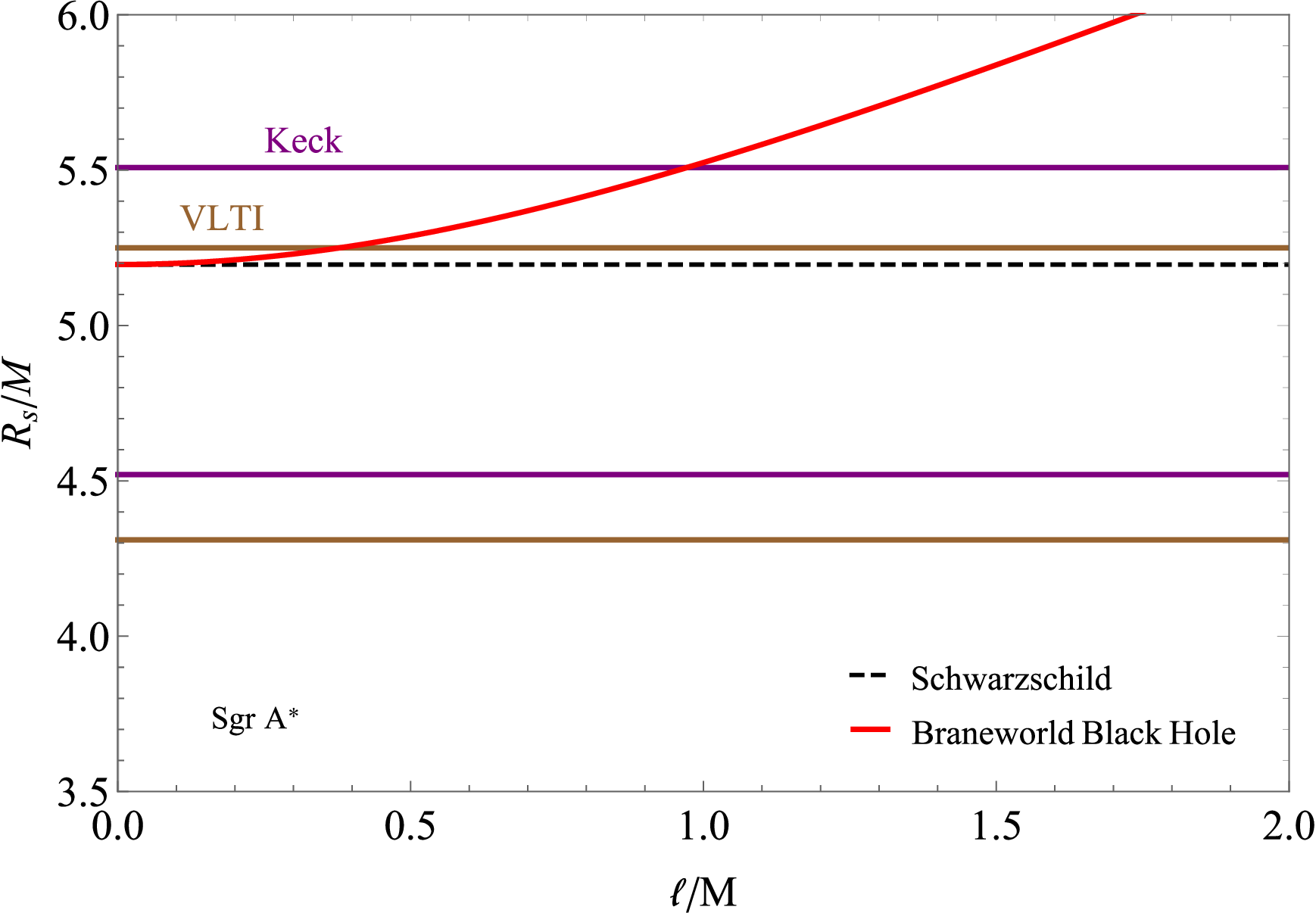}

}

\caption{\label{sha}{\footnotesize{}{}{}{}{}(a) Shows the changes
in the behavior of the shadow radius with the variation of the $\mathrm{AdS_{5}}$
radius $\ell$. In turn, (b) we found bounds from
the direct comparison between the results obtained by our theoretical
analysis with the EHT experimental data.}}
\end{figure}

In our analysis, we aim to obtain bounds on the $\mathrm{AdS_{5}}$
curvature radius. Therefore, the observed angular
radius---which has ringlike features in the horizon-scale EHT image
of Sgr A{*}---must be compared to the theoretically determined
angular radius of the black hole shadow. From the precise determination
of both mass and distance of Sgr A{*}, the central supermassive black
hole of our galaxy, we can verify the accordance, within the uncertainty
allowed, between the predicted theoretical results by this study and
the empirical data found by the EHT observations \citep{akiyama2022millimeter}.
On its turn, the mass and distance of Sgr A{*} have been studied over
the past few decades by exploring the dynamics of star clusters and,
in particular, the movement of stars close to the center of the galaxy.
According to Ref. \citep{EventHorizonTelescope:2022xqj},
the shadow radius measurements are restricted to the following region
determined by the estimates based on the Keck and Very Large Telescope Interferometer (VLTI) instruments,
within $1\sigma$ confidence interval, respectively, 
\begin{eqnarray}
4.5\lesssim R_{s}/M\lesssim5.5,\quad\text{and}\quad4.3\lesssim R_{s}/M\lesssim5.3.
\end{eqnarray}
In Fig. \ref{shadow}, we show the expected deviations in the behavior
of the shadow radius owing to the extra dimension. As we see, the
increase in the $\mathrm{AdS_{5}}$ curvature radius $\ell$ results
in a consequent increase in the radius of the black hole shadow. In
our analysis, we have considered the mass from Keck- and VLTI-based
estimates \citep{vagnozzi2022horizon}. Therefore, from the empirical data obtained by observations
of EHT, we can impose the upper limits on curvature radius at the $68\%$
confidence level (CL), such that
\begin{equation}
\ell_{\left(\text{Keck}\right)}\lesssim5.7\times10^{9}\,\mathrm{m}\simeq3.8\times10^{-2}\,\mathrm{AU},\quad\text{and}\quad\ell_{\left(\text{VLTI}\right)}\lesssim2.4\times10^{9}\,\mathrm{m}\simeq1.6\times10^{-2}\,\mathrm{AU},
\end{equation}
where $1\mathrm{AU}$ is one astronomical unit. These results are
shown in Fig. \ref{Shcomp}.

At this point, we can compare our results with another which investigates
similar physical systems. In this case, the absence of deviations
on observation of the highly circular shadow of black hole M87{*}
provides constraints on the $\mathrm{AdS_{5}}$ radius so that $\ell\lesssim170\,\mathrm{AU}$
\citep{Vagnozzi:2019apd}. In turn, the first bounds obtained on the
extra dimensions by the study of the time lag between the detection
of gravitational waves (GW) and electromagnetic (EM) signals from
the merging of a binary system detected by the event GW170817 have
established at $68\%$ CL an upper limit of $\ell\lesssim0.535\,\mathrm{Mpc}\simeq1.10\times10^{11}\,\mathrm{AU}$
\citep{Visinelli:2017bny}, whereas our analysis found $\ell_{\left(\text{Keck}\right)}\lesssim3.8\times10^{-2}\,\mathrm{AU}\left(\ell_{\left(\text{VLTI}\right)}\lesssim1.6\times10^{-2}\,\mathrm{AU}\right)$
for the same confidence interval. It is worth mentioning that, as
we have seen, our study yields more stringent constraints than those found
by similar physical systems, according to the best of our knowledge.
Finally, although we obtained weaker constraints compared to results
obtained from sub-mm scale gravity experiments, this approach has
set independent limits and can aid the investigation of the geometric
structure of the Universe.

\section{Final Remarks \label{S4}}

Recently, higher-dimensional theories have drawn increasing attention.
As has been pointed out, the seek for evidence of large extra dimensions
has been performed in several experimental researches, making its
detection a challenging problem in itself. By investigating
the null geodesic curves, we find theoretical prediction for the black
hole shadow radius, aiming to search for traces of extra dimensions.
From recent measurements of the shadow image of Sgr A{*}---a supermassive
black hole in the center of our Galaxy---we estimate deviations
owing to an extra dimension in the RSII scenario on the theoretical
shadow radius predicted. Given the inaccuracy involving the measured
spin of Sgr A{*}, which may even be zero but still can provide compatible
results with recent measures \citep{Fragione:2020khu}, we explore
the possibility of describing them as a nonrotating RSII black hole.
In this case, we have considered the Garriga-Tanaka solution that
takes a classical approach to linearized braneworld gravity and describes
a black hole solution in the weak field limit.

As we saw, the curvature radius of the $\mathrm{AdS_{5}}$ spacetime
is positively correlated to the shadow radius so that the increasing/decreasing
in one leads to correspondingly increasing/decreasing in the other.
Furthermore, we found that $\text{\ensuremath{\ell=0}}$, i.e., in
the absence of new physics coming from extra dimension, one recovers
the standard estimates expected for the Schwarzschild background.
From recent observational data of the shadow of Sgr A{*} released
by the EHT collaboration, one may impose constraints on the curvature
radius of $\mathrm{AdS_{5}}$ spacetime. Therefore,
by considering the Keck and VLTI estimates, we find more stringent
limits on the $\mathrm{AdS_{5}}$ curvature radius,
$\ell_{\left(\text{Keck}\right)}\lesssim3.8\times10^{-2}\,\mathrm{AU}$
and $\ell_{\left(\text{VLTI}\right)}\lesssim1.6\times10^{-2}\,\mathrm{AU}$
(at $68\%$ CL), than previously obtained for similar physical systems,
to the best of our knowledge. Besides, a nonzero predicted curvature
radius leads to agreement with the observed shadow radius at the $68\%$
confidence interval.

We would like to highlight that, although our analysis provides weaker
bounds than those obtained from collider experiments or sub-mm precision
gravity tests, which have set limits on the non-Newtonian force, in
this study, we have provided conservative bounds since we investigated the shadow radius in a nonrotating RSII-type black
hole background. Further, the bounds we reported were obtained independently and in a manner complementary to the previously mentioned studies. 
Finally, the newly begun era of observational astronomy opens several avenues for future research that might be explored for testing fundamental physics---and so getting new constraints on the $\mathrm{AdS_{5}}$ curvature radius, for instance---or even searching eventual deviations on GR, shedding light into seeking physics beyond the SM.

\begin{acknowledgments}
We thank the referee for the valuable comments and suggestions that have enhanced this work. We would like to thank CNPq, CAPES, and CNPq/PRONEX/FAPESQ-PB (Grant
No. $165/2018$), for partial financial support. F.A.B. acknowledges
support from CNPq (Grant No. 309092/2022-1). A.S.L. acknowledges support
from CAPES (Grant No. 88887.800922/2023-00). J.A.V.C. would like to
thank FAPESQ-PB/CNPq (Grant No. 077/2022) for financial support. 
\end{acknowledgments}

\bibliographystyle{ieeetr}

\begin{thebibliography}{99}
\bibitem{Nordstrom:1914ejq} G.~Nordstrom, \"Uber die M\"oglichkeit das elektromagnetische Feld und das Gravitationsfeld zu vereinigen,
{ Phys. Z.} \textbf{15}, 504 (1914).

\bibitem{Kaluza:1921tu} T.~Kaluza,
``Zum Unit\"atsproblem der Physik,''
Sitzungsber. Preuss. Akad. Wiss. Berlin (Math. Phys.) \textbf{1921}, 966 (1921).

\bibitem{Klein:1926tv} O.~Klein,
``Quantum Theory and Five-Dimensional Theory of Relativity,''
Z. Phys. \textbf{37}, 895 (1926).

\bibitem{Antoniadis:1998ig} I.~Antoniadis, N.~Arkani-Hamed, S.~Dimopoulos and G.~R.~Dvali,
New dimensions at a millimeter to a Fermi and superstrings at a TeV,
Phys. Lett. B \textbf{436}, 257 (1998).

\bibitem{Arkani-Hamed:1998jmv} N.~Arkani-Hamed, S.~Dimopoulos and G.~R.~Dvali,
The Hierarchy problem and new dimensions at a millimeter,
Phys. Lett. B \textbf{429}, 263 (1998).

\bibitem{Arkani-Hamed:1998sfv} N.~Arkani-Hamed, S.~Dimopoulos and G.~R.~Dvali,
Phenomenology, astrophysics and cosmology of theories with submillimeter dimensions and TeV scale quantum gravity,
Phys. Rev. D \textbf{59}, 086004 (1999).

\bibitem{Randall:1999ee} L.~Randall and R.~Sundrum,
A Large mass hierarchy from a small extra dimension,
Phys. Rev. Lett. \textbf{83}, 3370 (1999).

\bibitem{Randall:1999vf} L.~Randall and R.~Sundrum,
An Alternative to compactification,
Phys. Rev. Lett. \textbf{83}, 4690 (1999).

\bibitem{ParticleDataGroup:2022pth} R.~L.~Workman {\em et~al.},
{Review of Particle Physics}, Prog. Theor. Exp. Phys. \textbf{2022}, 083C01 (2022).

\bibitem{ATLAS:2021kxv} G.~Aad \textit{et al.} [ATLAS],
Search for new phenomena in events with an energetic jet and missing transverse momentum in $pp$ collisions at $\sqrt {s}$ =13  TeV with the ATLAS detector,
Phys. Rev. D \textbf{103}, 112006 (2021).

\bibitem{Hannestad:2003yd} S.~Hannestad and G.~G.~Raffelt, {Supernova
and neutron star limits on large extra dimensions reexamined}, Phys. Rev. D \textbf{67}, 125008 (2003); \textbf{69}, 029901(E) (2004).

\bibitem{Adelberger:2009zz} E.~G.~Adelberger, J.~H.~Gundlach, B.~R.~Heckel, S.~Hoedl, and S.~Schlamminger, {Torsion balance experiments:
A low-energy frontier of particle physics}, Prog. Part. Nucl. Phys. \textbf{62}, 102 (2009).

\bibitem{Murata:2014nra} J.~Murata and S.~Tanaka, {A review
of short-range gravity experiments in the LHC era},  Classical Quantum Gravity \textbf{32}, 033001 (2015).

\bibitem{Dahia:2015xxa} F.~Dahia and A.~S.~Lemos, {Constraints
on extra dimensions from atomic spectroscopy}, Phys. Rev. D \textbf{94}, 084033 (2016).

\bibitem{Dahia:2015bza} F.~Dahia and A.~S.~Lemos, {Is the proton
radius puzzle evidence of extra dimensions?}, Eur. Phys. J. C \textbf{76}, 435 (2016).

\bibitem{Dahia:2017jya} F.~Dahia, E.~Maciel, and A.~S.~Lemos, {Rydberg states of hydrogenlike ions in a braneworld model}, Eur. Phys. J. C \textbf{78}, 526 (2018).

\bibitem{Lemos:2019rhj} A.~S.~Lemos, G.~C.~Luna, E.~Maciel, and
F.~Dahia, {Spectroscopic tests for short-range modifications
of Newtonian and post-Newtonian potentials}, Classical Quantum Gravity \textbf{36}, 245021 (2019).

\bibitem{Lemos:2021xpi} A.~S.~Lemos, {Submillimeter constraints
for non-Newtonian gravity from spectroscopy}, Europhys. Lett. \textbf{135}, 11001 (2021).

\bibitem{Luo:2006ad} F.~Luo and H.~Liu, {Exploring extra dimensions
in spectroscopy experiments}, Chin. Phys. Lett. \textbf{23}, 2903 (2006).

\bibitem{Zhi-gang:2007swh} L.~Zhi-gang, W.-T.~Ni, and A.~P.~Paton,
{Extra dimensions and atomic transition frequencies}, Chin. Phys. B \textbf{17}, 70 (2008).

\bibitem{Salumbides:2015qwa} E.~J.~Salumbides, A.~N.~Schellekens,
B.~Gato-Rivera, and W.~Ubachs, {Constraints on extra dimensions
from precision molecular spectroscopy}, New J. Phys. \textbf{17}, 033015 (2015).

\bibitem{Lemos:2022vhw}A.~S.~Lemos and F.~A.~Brito, {Quantum gravity constraints on fine structure constant from GUP
in braneworlds}, Eur. Phys. 433 J. C \textbf{83}, 362 (2023).

\bibitem{CMS:2018ozv} A.~M.~Sirunyan {\em et~al.}, {Search
for black holes and sphalerons in high-multiplicity final states in
proton-proton collisions at $\sqrt{s}=13$ TeV}, J. High Energy Phys. \textbf{11}, 042 (2018).

\bibitem{ATLAS:2015yln} G.~Aad {\em et~al.}, Search for strong gravity in multijet final states produced in pp collisions at $\sqrt{s} =$ 13 TeV using the ATLAS detector at the LHC, J. High Energy Phys. \textbf{03}, 026 (2016).

\bibitem{Emparan:2002jp} R.~Emparan, J.~Garcia-Bellido, and N.~Kaloper,
{Black hole astrophysics in AdS brane worlds}, J. High Energy Phys.
\textbf{01}, 079 (2003).

\bibitem{Bohra:2023vls} S.~S.~Bohra, S.~Sarkar, and A.~A.~Sen,
{Gravitational atoms in the braneworld scenario},  Phys. Rev. D \textbf{109}, 104021 (2024).

\bibitem{Chrysostomou:2022evl} A.~Chrysostomou, A.~Cornell, A.~Deandrea,
E.~Ligout, and D.~Tsimpis, {Black holes and nilmanifolds: quasinormal
modes as the fingerprints of extra dimensions?}, Eur. Phys. J. C \textbf{83}, 325 (2023).

\bibitem{McWilliams:2009ym} S.~T.~McWilliams, {Constraining
the braneworld with gravitational wave observations}, Phys. Rev. Lett. \textbf{104}, 141601 (2010).

\bibitem{Vagnozzi:2019apd} S.~Vagnozzi and L.~Visinelli, {Hunting
for extra dimensions in the shadow of M87{*}}, Phys. Rev. D \textbf{100}, 024020 (2019).

\bibitem{Visinelli:2017bny} L.~Visinelli, N.~Bolis, and S.~Vagnozzi,
{Brane-world extra dimensions in light of GW170817}, Phys. Rev. D \textbf{97}, 064039 (2018).

\bibitem{Mishra:2021waw} A.~K.~Mishra, A.~Ghosh, and S.~Chakraborty, {Constraining extra dimensions using observations of black hole
quasi-normal modes}, Eur. Phys. J. C \textbf{82}, 820 (2022).

\bibitem{Hou2021}Y.~Hou, M.~Guo and B.~Chen, Revisiting the
shadow of braneworld black holes,  Phys. Rev. D \textbf{104}, 024001 (2021).

\bibitem{Chakraborty:2017qve}
S.~Chakraborty, K.~Chakravarti, S.~Bose and S.~SenGupta,
Signatures of extra dimensions in gravitational waves from black hole quasinormal modes,
Phys. Rev. D \textbf{97}, 104053 (2018).

\bibitem{Dey:2020pth}
R.~Dey, S.~Biswas and S.~Chakraborty,
Ergoregion instability and echoes for braneworld black holes: Scalar, electromagnetic, and gravitational perturbations,
Phys. Rev. D \textbf{103}, 084019 (2021).

\bibitem{Hashemi:2019jlt}
S.~S.~Hashemi, M.~Kord~Zangeneh and M.~Faizal,
Charged scalar quasi-normal modes for higher-dimensional Born\textendash{}Infeld dilatonic black holes with Lifshitz scaling,
Eur. Phys. J. C \textbf{80}, 111 (2020).

\bibitem{Anacleto:2023ntm}
M.~A.~Anacleto, J.~A.~V.~Campos, F.~A.~Brito, E.~Maciel and E.~Passos,
Scattering and absorption by extra-dimensional black holes with GUP,
Nucl. Phys. B \textbf{1006}, 116617 (2024).

\bibitem{Cruz:2020emz}
M.~B.~Cruz, F.~A.~Brito and C.~A.~S.~Silva,
Polar gravitational perturbations and quasinormal modes of a loop quantum gravity black hole,
Phys. Rev. D \textbf{102}, 044063 (2020).

\bibitem{Cardoso:2003cj}
V.~Cardoso, R.~Konoplya and J.~P.~S.~Lemos,
Quasinormal frequencies of Schwarzschild black holes in anti-de Sitter space-times: A Complete study on the asymptotic behavior,
Phys. Rev. D \textbf{68}, 044024 (2003).

\bibitem{Konoplya:2011qq}
R.~A.~Konoplya and A.~Zhidenko,
Quasinormal modes of black holes: From astrophysics to string theory,
Rev. Mod. Phys. \textbf{83}, 793 (2011).

\bibitem{Konoplya:2003ii}
R.~A.~Konoplya,
Quasinormal behavior of the d-dimensional Schwarzschild black hole and higher order WKB approach,
Phys. Rev. D \textbf{68}, 024018 (2003).

\bibitem{Yagi:2016jml}
K.~Yagi and L.~C.~Stein,
{Black Hole Based Tests of General Relativity},  Classical Quantum Gravity \textbf{33}, 054001 (2016).

\bibitem{Goddi:2016qax}
C.~Goddi, H.~Falcke, M.~Kramer, L.~Rezzolla, C.~Brinkerink, T.~Bronzwaer, J.~R.~J.~Davelaar, R.~Deane, M.~D.~Laurentis and G.~Desvignes, \textit{et al.}
{BlackHoleCam: Fundamental physics of the galactic center},  Int. J. Mod. Phys. D \textbf{2}, 1730001 (2016).

\bibitem{EventHorizonTelescope:2019dse} K.~Akiyama {\em et~al.},
{First M87 Event Horizon Telescope Results. I. The Shadow of the
Supermassive Black Hole}, Astrophys. J. Lett. \textbf{875}, L1 (2019).

\bibitem{EventHorizonTelescope:2019ggy} K.~Akiyama {\em et~al.},
{First M87 Event Horizon Telescope Results. VI. The Shadow and
Mass of the Central Black Hole}, Astrophys. J. Lett. \textbf{875}, L6 (2019).

\bibitem{EventHorizonTelescope:2022xqj} K.~Akiyama {\em et~al.},
``{First Sagittarius A{*} Event Horizon Telescope Results. VI. Testing
the Black Hole Metric}, Astrophys. J. Lett. \textbf{930}, L17 (2022).

\bibitem{EventHorizonTelescope:2022wkp} K.~Akiyama {\em et~al.},
``{First Sagittarius A{*} Event Horizon Telescope Results. I. The
Shadow of the Supermassive Black Hole in the Center of the Milky Way}, Astrophys. J. Lett. \textbf{930}, L12 (2022).

\bibitem{vagnozzi2022horizon} S.~Vagnozzi, R.~Roy, Y.-D.~Tsai,
L.~Visinelli, M.~Afrin, A.~Allahyari, P.~Bambhaniya, D.~Dey,
S.~G. Ghosh, P.~S. Joshi, {\em et~al.}, {Horizon-scale tests
of gravity theories and fundamental physics from the Event Horizon
Telescope image of Sagittarius A}, Classical Quantum Gravity \textbf{40}, 165007 (2023).

\bibitem{Vagnozzi2022} S.~Vagnozzi and L.~Visinelli, Note on
Fundamental Physics Tests from Black Hole Imaging: Comment on Hunting
for Extra Dimensions in the Shadow of Sagittarius A{*}, Res. Notes AAS \textbf{6}, 106 (2022).

\bibitem{Garriga:1999yh} J.~Garriga and T.~Tanaka, {Gravity
in the brane world}, Phys. Rev. Lett. \textbf{84}, 2778 (2000).

\bibitem{Banerjee:2022jog} I.~Banerjee, S.~Chakraborty, and S.~SenGupta,
{Hunting extra dimensions in the shadow of Sgr A{*}}, Phys. Rev. D \textbf{106}, 084051 (2022).

\bibitem{Fragione:2020khu} G.~Fragione and A.~Loeb, {An upper
limit on the spin of SgrA$^{*}$ based on stellar orbits in its vicinity}, Astrophys. J. Lett. \textbf{901}, L32 (2020).

\bibitem{Anacleto:2021qoe}
M.~A.~Anacleto, J.~A.~V.~Campos, F.~A.~Brito and E.~Passos,
Quasinormal modes and shadow of a Schwarzschild black hole with GUP, Ann. Phys. (Amsterdam) \textbf{434}, 168662 (2021).

\bibitem{Campos:2021sff}
J.~A.~V.~Campos, M.~A.~Anacleto, F.~A.~Brito and E.~Passos,
Quasinormal modes and shadow of noncommutative black hole, Sci. Rep. \textbf{12}, 8516 (2022).

\bibitem{cardoso2009geodesic} V.~Cardoso, A.~S.~Miranda, E.~Berti,
H.~Witek, and V.~T.~Zanchin, Geodesic stability, lyapunov exponents,
and quasinormal modes, Phys. Rev. D \textbf{79}, 064016 (2009).

\bibitem{vazquez2003strong} S.~E.~Vazquez and E.~P.~Esteban, {Strong
field gravitational lensing by a Kerr black hole}, Nuovo Cimento Soc. Ital. Fis. B \textbf{119}, (2004).

\bibitem{akiyama2022millimeter} K.~Akiyama, J.~Kauffmann, L.~D.~Matthews, K.~Moriyama, S.~Koyama, and K.~Hada, Millimeter/submillimeter
VLBI with a next generation large radio telescope in the atacama desert, Galaxies \textbf{11}, 1 (2022).
\end{thebibliography}

\end{document}